\documentclass[12pt,a4paper]{article}
\usepackage[english]{babel}
\usepackage[latin1]{inputenc}
\usepackage[T1]{fontenc}
\usepackage{amsmath}
\usepackage{amsfonts}
\usepackage{amssymb}
\usepackage{braket}
\usepackage[dvips]{graphicx,color}
\usepackage{hyperref}
\begin{document}
\title{Dynamical Casimir effect from fermions in an oscillating bag in $1+1$ dimensions} 
\author{C.~D.~Fosco and
G.~Hansen\\
{\normalsize\it Centro At\'omico Bariloche and Instituto Balseiro}\\
{\normalsize\it Comisi\'on Nacional de Energ\'{\i}a At\'omica}\\
{\normalsize\it R8402AGP S.\ C.\ de Bariloche, Argentina.} }
\maketitle
\begin{abstract} 
We evaluate dissipative effects for a system consisting of a massive Dirac
	field confined between two walls, one of them oscillating, in $1+1$
	dimensions. In the model that we consider, a dimensionless
	parameter characterizing each wall is tuned so that bag-boundary
	conditions are attained for a particular value.  We present
	explicit results for the probability of creating a fermion pair, and relate the total probability to the imaginary
	part of the effective action.  
\end{abstract}
\maketitle
\section{Introduction}
Quantum Field Theory predicts many interesting effects in the presence of
nontrivial boundary conditions.  The best known example of this phenomenon
is the Casimir effect~\cite{Casimir:1948dh,Bordag:2001qi} which, in the
static case, manifests itself in forces due to a non-trivial dependence of
the vacuum energy on the geometry of the boundary. Although, in principle,
this effect is relevant for any kind of fluctuating field, the most
frequently studied case corresponds to an Abelian gauge field. This is
hardly surprising since, for the electromagnetic (EM) field, boundary
conditions can be controlled in a rather precise and straightforward way.
Nevertheless, fields other than the electromagnetic field have also been
studied, like in the fermionic fields describing quarks, since their vacuum
energies  play an important role in the bag model of
QCD~\cite{Chodos:1974pn}, where part of the mass of a baryon is due to the
Casimir energy of the fields which are affected by the (bag) boundary
conditions.  A more straightforward realization arises in the context of
Condensed Matter Physics, where Dirac fields  play a preeminent role,
specially in $1+1$ and $2+1$ dimensions~\cite{Fradkin}. 
Boundary conditions may, on the other hand, be also relevant due to the
existence of impurities, domain walls, etc.  

We are interested here in the dynamical Casimir effect, whereby a time
dependence of the boundary may induce the creation of particles of the
quantum field out of the vacuum. In~\cite{Mazzitelli:1986qk} this has been
studied, for a massless Dirac field in $1+1$ dimensions satisfying bag
conditions on two moving boundaries. For massive Dirac fields, higher
dimensions, and more general boundary condition, the imaginary part the
effective action for a single moving boundary has been evaluated
in~\cite{Fosco:2007nz}.  In this paper, we consider a massive Dirac field
coupled to two walls, one them moving, both imposing boundary conditions
which, for a particular value of a parameter describing the coupling of the
fermion to the wall, correspond to the vanishing of the component of the
current which is normal to the boundary: bag conditions.

The structure of this paper is as follows: in Section~\ref{sec:themodel} we
introduce the concepts and define the model that we study in the rest of
this work. Then, in Section~\ref{sec:transition}, we evaluate the
probability of pair creation from the vacuum, assuming a small oscillation
amplitude.  In Section~\ref{sec:eff}, we compare, and show the consistency
of the previous result with the one that one finds from the evaluation of
the imaginary part of the effective action. Finally, in Section \ref{sec:conc}
we present our conclusions.

\section{The model}\label{sec:themodel}
In the model that we consider, the (real-time) action ${\mathcal S}$, describing 
the fermionic field ($\psi$, ${\bar\psi}$) subjected to  boundary conditions, is:
\begin{equation}\label{eq:defsf}
{\mathcal S}({\bar\psi},\psi;V)\;=\;
\int d^2x \,{\bar\psi}(x) \, {\mathcal D} \, \psi(x)
\end{equation}
with 
\begin{equation}
{\mathcal D} \equiv i\not \! \partial - m - V(x) \;,
\end{equation}
where $m$ is the mass of the fermion field, and $V(x)$ will be used in
order to introduce the boundary conditions (see below).  In our
conventions, both $\hbar$ and the speed of light are equal to $1$, 
the spacetime coordinates are denoted by $x^\mu$, $\mu\,=\,
0,\,1$, $x^0 = t$, and the metric tensor is \mbox{$g_{\mu\nu} \equiv {\rm diag}(1,-1)$}.
Dirac's $\gamma$-matrices are chosen as follows:
\begin{equation}\label{eq:gamma_matrices}
\gamma^0 \,\equiv\, \sigma_1 \,=\, 
\left(
\begin{array}{cc}
	0 & 1 \\
	1 & 0
\end{array}
\right)
\;,\;\;
\gamma^1 \,\equiv\, i \, \sigma_3 \,=\, 
\left(
\begin{array}{cc}
i & 0 \\
0 & -i 
\end{array}
\right) \;,
\end{equation}
and 
\begin{equation}
\gamma^5 \,\equiv\, \gamma_5 \,\equiv\, \gamma^0 \gamma^1 \,=\, 
\sigma_2 \,=\,\left(
\begin{array}{cc}
0 & -i \\
i & 0
\end{array}
\right) \,,  
\end{equation}
with $\sigma_i$ ($i=1,\,2,\,3$) representing the usual Pauli's matrices.

Following the approach of~\cite{EfeCasFer,Ttira:2010rh}, we can impose
boundary conditions by a special choice of the `potential' $V$. Namely, $V$
has to be proportional to a $\delta$-function concentrated on the worldline
swept by the point where the condition is imposed. For example, for a
time-like curve ${\mathcal C}$, corresponding to the solution to the equation $F(x) = 0$,
the potential $V$ shall have the structure:
\begin{equation}\label{eq:defV}
	V(x) \;=\; g \, |N| \; \delta[F(x)] \;,
\end{equation}
where $|N| \equiv \sqrt{- N_\mu N^\mu}$,  $N_\mu \equiv \pm [\partial_\mu
F(x)]_{F=0}$, is defined on ${\mathcal C}$, and everywhere normal to it
(therefore space-like). There is a global sign ambiguity in $N_\mu$, which
corresponds to the two possible orientations of the normal to a curve. We
will fix it by setting it to point towards the interior of the region limited by
two curves.

When ${\mathcal C}$ is the union of disconnected curves, $V$ decomposes
into a sum of terms, one for each curve.
The factor $g$, on the other hand, is a constant.

We shall assume that there are two walls, i.e, two curves $L$ and $R$
(which eventually become boundaries in the bag limit). $L$ is static and
given by $x^1\,=\,0$, while the other, $R$, has the trajectory $x^1 = a +
\eta(x^0)$  ($\eta(x^0) \,>\,-a$). 

Applying the general structure of $V$ discussed above to the case at hand,
it will consist of two terms, namely, 
\begin{equation}\label{eq:V}
	V(x) \;=\; g_L \, \delta(x^1) \,+\, g_R \, \gamma^{-1}(\dot{\eta}(x^0)) \,
\delta(x^1 - q(x^0)) \;,
\end{equation}
where $\gamma(u) \,\equiv \,1/\sqrt{1 - u^2}$ is the Lorentz factor. 

Here, $g_L$ and $g_R$ are constants which, in order to enforce bag boundary
conditions, have to equal $2$ (see~\cite{EfeCasFer}). Different values
produce `imperfect' boundary conditions, in the sense that some current may
escape the cavity.  We recall that the general form of the bag boundary
conditions
\begin{equation}\label{eq:bbc}
	 \big(e^{i \theta \gamma_5} \;+\; i \, n^\mu \gamma_\mu \big)\, 
	\psi \Big|_{\mathcal C} \;=\; 0 \;, 
\end{equation}
where $\theta$ is a real parameter which can be chosen arbitrarily, and
$n^\mu \equiv \frac{N^\mu}{|N|}$. 
Note that, as usual, the boundary condition is assumed to be imposed on the
limit of the function on which it acts, when one approaches the curve from
the interior of the region delimited.

Since we are going to deal with the region limited between $L$ and $R$, on
$L$, $n^\mu = \delta^\mu_1$, while on $R$, $n^\mu(x^0) \,= - \gamma(\dot{q}) \,
(\delta^\mu_0 \dot{q} + \delta^\mu_1)$ \;.

To see the kind of boundary condition due to  a singular term like the one
we are considering, let us observe  what happens for a singularity of strength
$g$ at $x^1=0$. We see from the Dirac equation, after integrating along a
spatial path  from $x^1 = -\epsilon$ and $x^1 = \epsilon$, that the
presence of the singular term introduces a discontinuity in $\psi$.
Therefore, following~\cite{Fosco:2007ry}, we replace the integral of the
$\delta$-function times $\psi$ by the average of the two lateral limits:
\begin{equation}
i \gamma^1 ( \psi(\epsilon) - \psi(-\epsilon) ) \,-\,  \frac{g}{2} 
	\big( \psi(\epsilon) + \psi(-\epsilon) \big)  \;=\;0 \;,
\end{equation}
where we have omitted writing the temporal arguments, which are the same
in all the terms.

Setting $g=2$, and introducing the orthogonal projectors: ${\mathcal
P}^{\pm} \equiv \frac{1 \pm i \gamma^1}{2}$, this is equivalent to:
\begin{equation}
{\mathcal P}^+  \psi(-\epsilon)  \;=\; - \, 
{\mathcal P}^-  \psi(\epsilon)  \;,
\end{equation}
and, therefore,
\begin{equation}
{\mathcal P}^+  \psi(-\epsilon)  \;=\;  0 \;\;,\;\;\;\;
{\mathcal P}^-  \psi(\epsilon)  \;=\;0 \;.
\end{equation}
The second equation is the bag boundary condition one has on the field
on $L$ (assuming $\theta = 0$), assuming the interior of the cavity is
between $L$ and $R$.

This formal argument will be seen to hold true in more concrete terms, in
Section~\ref{sec:eff}, when evaluating different terms in the perturbative
expansion of the effective action $\Gamma(q)$, that results by functional
integrating out the Dirac field in the vacuum to vacuum transition
amplitude:
\begin{equation}\label{eq:defgef}
	e^{i \Gamma(q)} \;=\; \frac{\int {\mathcal D}\psi {\mathcal
	D}{\bar\psi} \, e^{i {\mathcal S}({\bar\psi},\psi;V)}}{\int 
	{\mathcal D}\psi {\mathcal D}{\bar\psi} \, 
	e^{i {\mathcal S}({\bar\psi},\psi;V_0)}} \;.
\end{equation}
Here $V$ is as defined in (\ref{eq:V}), and we have introduced $V_0$, the
function $V$ corresponding to $q \equiv a$, where $a$ is a positive constant. 
The denominator thus incorporates the static Casimir effect, which has  been evaluated for
this case~\cite{EfeCasFer}, where it has been shown that it properly
reproduces the fermionic Casimir force for bag boundary conditions, when
$g_L = g_R = 2$. For different values of $g$, the strength of the
interaction is weaker.

\section{Pair creation}\label{sec:transition}
We evaluate there the probability of pair creation out of the vacuum, due
to the motion of one of the walls, which acts as an `external source' injecting
energy into the system.
We will consider motions of the $R$ wall which are parametrized by means of
a function $\eta(x^0)$, which measures the departure of $R$ from its
equilibrium, time average position
$a>0$, namely, 
\begin{equation}
q(x^0) = a + \eta(x^0) \;.
\end{equation}
The object we study is the $S$-matrix; more specifically, matrix elements
of the $T$-matrix which describes the non trivial part of the evolution:
\begin{equation}
	S \;=\; 1 \,+\, i \, T \;.
\end{equation}
For the perturbative evaluation of those matrix elements, we will make
use of the interaction representation. Note, however, that bag conditions
correspond to $g=2$, thus, an expansion in powers of $g$ is impossible.  
We can, however, use a reliable expansion
which captures interesting physics, by taking as unperturbed system the one
corresponding to two static boundaries (separated by a distance $a$) and
the difference between the real action and the unperturbed one as
perturbation. This may be justified if one assumes, as we do, that the
departure $\eta$ is sufficiently small.
Thus, the action is split up as follows:
\begin{equation}\label{eq:split1}
{\mathcal S}\;=\; {\mathcal S}_0 \,+\, {\mathcal S}_I
\end{equation}
with:
\begin{equation}\label{eq:split2}
{\mathcal S}_0 \;\equiv \; {\mathcal S}({\bar\psi},\psi;V_0) \;\;,\;\;\;
V_0(x) \;=\; 2 \, \delta(x^1) \,+\, 2 \, \delta(x^1 - a) \;,
\end{equation}
and
\begin{equation}
{\mathcal S}_I \;\equiv\; - \int d^2x \, {\bar\psi}(x) \varphi(x) \psi(x) 
\;\;,\;\;\; \varphi(x) \;\equiv\; V(x) - V_0(x) \;.
\end{equation}
In ${\mathcal S}_I$, the fields are in the interaction picture, so that
their time evolution is dictated by the free Hamiltonian, which corresponds
to the potential $V_0$: static walls (at a distance $a$).

Then, we evaluate the transition amplitudes that result by expanding $T$ in
powers of ${\mathcal S}_I$,  for small departures $\eta$.
Up to the second order in $\eta$, we see that 
$\varphi = \varphi^{(1)} + \varphi^{(2)} + \ldots$, with
\begin{align}
\varphi^{(1)}(x) \;&=\;- 2 \, \delta'(x^1 -  a) \, \eta(x^0) \\ 
	\varphi^{(2)}(x) \;&=\;\, 2 \, \left[ \delta''(x^1 - a)
	\, \big(\eta(x^0)\big)^2 \,
	+\,\delta(x^1 - a) \,  \big(\dot{\eta}(x^0)\big)^2 \right]  \;, 
\end{align}
where the prime denotes differentiation with respect to $x^1$.

Let us now evaluate, to the lowest non-trivial order in $\eta$, the transition
amplitudes and transition probabilities (the latter will be of the second
order in $\eta$), assuming the initial state to be the vacuum of the
unperturbed system.
To the first order in $\eta$, the transition amplitude from $\ket{i}$ to $\ket{f}$
is:
\begin{align}\label{eq:Tfi}
T_{fi}^{(1)} \;=\; \langle f |{\mathcal S}_I | i \rangle &=\;
-\int d^2x \, \varphi^{(1)}(x) \, \langle f |\bar{\psi}(x)\psi(x) |i
\rangle \nonumber\\
&=\; -\int d^2x \, \varphi^{(1)}(x) \braket{f|: \bar{\psi}(x)\psi(x) :|i}
	\;.
\end{align}
The normal ordering above is justified as follows: using Wick's theorem in
${\mathcal S}_I$,
\begin{align}
{\mathcal S}_I &=\; \int d^2x \, \varphi^{(1)}(x) \, \bar{\psi}(x)\psi(x) 
\,=\, \int d^2x \, \varphi^{(1)}(x) \, 
\Big( :\bar{\psi}(x)\psi(x): \nonumber\\
& -\, {\rm Tr}[S_F(x,x)] \Big) \, 
\end{align}
where $S_F$ is the fermion propagator in the presence of the static boundaries.
Now, the term involving $S_F$ vanishes. Indeed, this object is invariant
under time translations: $S_F(x^0, x^1; x'^0,x'^1)=S_F(x^0-x'^0;x^1,x^1)$.
Thus,
\begin{align}
\int d^2x \,\varphi^{(1)}(x) \,{\rm tr}[S_F(x,x)] \,=\,   
\int d^2x \,\varphi^{(1)}(x) \,{\rm tr}[S_F(0;x^1,x^1)] \nonumber\\
	=\, - 2 \, \Big( \int dx^0 \,\eta(x^0) \Big)\; 
\int dx^1 \delta'(x^1-a)  {\rm tr}[S_F(0;x^1,x^1)] \nonumber\\
=\, - 2 \,  \int dx^0 \,\langle \eta \rangle \; 
\int dx^1 \delta'(x^1-a)  {\rm tr}[S_F(0;x^1,x^1)] \,=\, 0 \;,
\end{align}
where $\langle \eta \rangle$ is the time average  of $\eta(x^0)$ which, by assumption, vanishes, since it is the {\em departure\/}
with respect to the average position $a$.
 On the other hand, note that $\langle \eta \rangle$ is multiplied by a factor which is divergent. Indeed, the coincidence limit picks up a logarithmic divergence, so that the UV behavior of that term is:
\begin{equation}
\int d^2x \,\varphi^{(1)}(x) \,{\rm tr}[S_F(x,x)] \,\sim \,   
 - 2 \,  \int dx^0 \,\langle\eta\rangle \; 
\frac{m}{a} \, \log(\frac{\Lambda}{m}) \;,
\end{equation}
where $\Lambda$ is an UV cutoff. The physical meaning of such a term in the action, is a divergent contribution to the static energy, not to the dynamical process we want to study and therefore one could have defined the theory with the normal ordering from the very beginning without affecting transition probabilities. Also, note that, if $\langle \eta \rangle$ were a non-vanishing constant, one could still absorbe that term, by a redefinition of $a$: $a \to a + \langle \eta \rangle$ in ${\mathcal S}_0$ and expanding to first order in $\langle \eta \rangle$, as it should be.

We want to study  particle production out of the vacuum, so that 
the initial state is $\ket{i} \equiv \ket{0}$; on the other hand, to this
order, the only kind of final state allowed contains a fermion anti-fermion pair. 
Note that this pair will not correspond to free space particles, rather, to
states contained in the bag, which are the eigenstates of the unperturbed
Hamiltonian.  They will be of the form $\ket{f} \equiv b_n^\dagger d_l^\dagger \ket{0}$,
with $b_n^\dagger$ and $d_l^\dagger$ being creation operators of 
fermions and anti-fermions, respectively. They are labelled by discrete
indices, $n$ and $l$, which correspond to spatial momenta when $a \to
\infty$. 
Indeed, a mode-expansion of the field operator (interaction picture) may be
constructed as follows:
\begin{equation}\label{eq:modeexp}
	\psi(x) \equiv \sum_{n} \left[ b_n \, e^{-i E_n x^0} u_n(x^1)
	+ \, d_n^\dagger \, e^{i E_n x^0} v_n(x^1)\right],
\end{equation}
where $u_n(x^1) \equiv \psi_{n,+}(x^1)$ and $v_n(x^1) \equiv \psi_{n,-
}(x^1)$, with $\psi_{n,\pm}$ are normalized solutions of Dirac equation
with bag boundary conditions (\ref{eq:bbc}):
\begin{equation}\label{eq:solutions}
	\begin{aligned}
		&\psi_{n,\pm}(x^1) = N_n
		\begin{pmatrix}
			\pm \frac{E_n}{p_n} \, \sin(p_n x^1) \\ 
			\cos(p_n x^1) + \frac{m}{p_n} \, \sin(p_n x^1)
		\end{pmatrix},\\
		&N_n \equiv \sqrt{2} \, p_n^2 \, [p_n^2 \, (m + 2 a E_n^2)
		+ m E_n^2 \sin^2(p_n a)]^{-1/2},
	\end{aligned}
\end{equation}
where $E_n \equiv \sqrt{p_n^2 + m^2}$ and the values of $p_n$ are
determined by a transcendental equation. In terms of the dimensionless
quantities $\rho_n \equiv p_n a$ and $\mu \equiv m a$, the energies may
also be rendered in units of $1/a$, introducing dimensionless energies
$\epsilon_n$:
$E_n = \frac{1}{a} \epsilon_n$,  $\epsilon_n =\sqrt{ \rho_n^2 + \mu^2}$,
while the transcendental equation is:
\begin{equation}
	\mu \, {\rm sinc} \, \rho_n  \, + \,  \cos \rho_n  \;=\; 0,
\end{equation}
with the ${\rm sinc}$ function defined as ${\rm sinc} \, x  =
\frac{\sin x}{x}$.  This yields a discrete
spectrum~\cite{Mamaev:1980,Bordag:2009}.
In the massless ($\mu \to 0$) limit, this spectrum is simply 
$p_n = (n + \frac{1}{2})\frac{\pi}{a} $ with $n = 0,1,\ldots$, and  
energies $\epsilon_n = \rho_n$.
In the opposite regime, $\mu \gg 1$, the spectrum  is in turn determined
by the zeros of the ${\rm sinc}$ function, namely: $p_n  = \frac{n
\pi}{a}$, $n = 1, 2, \ldots$. Note that the lowest energy is, in this
limit, the mass of the fermions.

Taking into account the mode expansion above, the transition amplitude for this
kind of process becomes: 
\begin{equation}\label{eq:tfi}
	T_{fi}^{(1)} \;\equiv \; T_{nl}\;= \; - 2 \, \tilde{\eta}(E_n + E_l)  \,
	\left(\bar{u}_n(x^1)v_l(x^1)\right)^\prime \Big|_{x^1 = a},
\end{equation}
where the Fourier transform of the departure is defined as: $\tilde{\eta}(\nu)
\equiv \int dx^0 e^{i \nu x^0} \eta(x^0)$. 
Using the explicit form of the eigenstates $u$ and $v$, we may write:
\begin{equation}
T_{nl} \;= \;  -\frac{4}{a^2} \, \tilde{\eta}(E_n + E_l) 
\; \xi_n \, \xi_l \; (\epsilon_n - \epsilon_l)   \;,
\end{equation}
with
\begin{equation}
\xi_n \;\equiv\; \frac{\epsilon_n {\rm sin}(\rho_n)}{\sqrt{2 \epsilon_n^2 +
	\mu \big(1 + \epsilon_n^2 \, {\rm sinc}^2\rho_n \big)}} \;.
\end{equation}
From the form of the matrix element of $T$ it is clear that, for the
transition to be possible, the energies $E_n$ and $E_l$ must be different.

We then write the probability of creation of
a specific pair in a spectral form, as follows:
\begin{equation}
	P_{nl}^{(1)} \;= \; \int \frac{d\nu}{2\pi} \, \gamma_{nl}(\nu) \, 
\big| \tilde{\eta}(\nu) \big|^2 \;.
\end{equation}
where:
\begin{equation}
	\gamma_{nl}(\nu) \;=\; \frac{32\pi}{a^4} \,  \delta[\nu- (E_n + E_l)]  \;
	\big[\xi_n \, \xi_l \; (\epsilon_n - \epsilon_l) \big]^2  \;.
\end{equation}

For strictly massless fermions, this becomes:
\begin{equation}
\gamma_{nl}(\nu) \;=\; \frac{8\pi^3}{a^4} \, \delta[\nu- \frac{(n +
	l + 1)\pi}{a}]  \; (n-l)^2  \;. 
\end{equation}
In this case, the frequency threshold $\nu_0$ required to produce a pair is then
given by considering $n=0$ and $l=1$. Thus $\nu_0 = \frac{\pi}{2 a} + 
\frac{3\pi}{2 a} = \frac{2\pi}{a}$. 

In the $\mu \to \infty$ limit, on the other hand, the probability is of
course $0$, since $\sin \rho_n$ (and therefore $\xi_n$) vanishes.

Finally, the {\em total\/} probability of pair creation $P$ is obtained by
summing over all values of $n$ and $l$ which give non-vanishing contributions.
\begin{equation}
	P \; = \; \sum_{n,l} P_{nl}^{(1)} \;= \; \int \frac{d\nu}{2\pi} \, \gamma(\nu) \, 
\big| \tilde{\eta}(\nu) \big|^2 \;,
\end{equation}
with
\begin{equation}
\gamma(\nu) \;=\;  \sum_{n,l} \gamma_{nl}(\nu) \;.
\end{equation}
In particular, for the massless case, we may write:
\begin{equation}
\gamma(\nu) \;=\;  \frac{8\pi^3}{a^4} \;
	\sum_{k=1}^\infty \, \delta[\nu- \frac{(k + 1)\pi}{a}]  \; f(k) \;,
\end{equation}
with
\begin{equation}
f(k) \;=\;\sum_{j=0}^k (2j - k)^2 \;, 
\end{equation}
where we have taken into account the fact that the minimum value of the
frequency threshold is $\frac{2\pi}{a}$.

The sum over $j$ may be explicitly evaluated, leading to the result:
\begin{equation}
	f(k) = \frac{1}{3} \, k \, \left(k + 1\right) \left(k + 2\right).
\end{equation}

\section{Imaginary part of the effective action}\label{sec:eff}
Let us here consider the (in-out) effective action $\Gamma$, in order to
check the consistency of its imaginary part with the pair creation probability
just derived. 
$\Gamma$ may be written as a functional trace, in terms of the
fermion propagator $S_F$  in the presence of the static boundaries, 
and of $\varphi \equiv V - V_0$, as follows:
\begin{equation}\label{eq:fdet1}
	\Gamma(q) \;=\; - i \, {\rm Tr}\, {\rm log} \left( 1 \,+\, i \, S_F
	\, \varphi \right) \;.
\end{equation}
In our conventions, $S_F$ is determined by $\big[i \not\!\partial - m -
V_0(x)\big] S_F = i I $, where $I$ is the identity operator in both
functional and spinorial spaces, also refer to its kernel. Since $V_0$ is time-independent,
we will use its Fourier transform 
\begin{equation}
	S_F(x^0-y^0; x^1, y^1) \,=\, \int \frac{d\omega}{2\pi} \, e^{-i
	\omega (x^0-y^0)} \, \widetilde{S}_F(\omega; x^1, y^1) \;. 
\end{equation}

Expanding for small departures, as in the previous Section, 
\begin{equation}
\Gamma \;=\;	\Gamma^{(0)}\,+\,\Gamma^{(1)} \,+\,\Gamma^{(2)}\,+\,\ldots
\end{equation}
where the index denotes the order of the term. 
It is rather straightforward to see that, since in our definition of
$\Gamma$ the static contribution is subtracted, then $\Gamma^{(0)}=0$.
Besides, from the assumption that the time average position of $R$ is $a$,
it follows that also the first order term vanishes. We thus only need to
evaluate $\Gamma^{(2)}$. 
On the other hand, we see that in its second-order term
there will be two qualitatively different contributions:
\begin{equation}
	\Gamma^{(2)} \;=\; \Gamma^{(2,1)} \,+\, \Gamma^{(2,2)} \;, 
\end{equation}
with
\begin{equation}
\label{eq:gammaexp2}
\Gamma^{(2,1)} \;=\;  {\rm Tr}\, \left( S_F \,\varphi^{(2)}\right)
	\;\;,\;\;\;\;
\Gamma^{(2,2)} \;=\; - \frac{i}{2} {\rm Tr}\, 
\left( S_F \,\varphi^{(1)}\, S_F \,\varphi^{(1)}\right) \;.
\end{equation}

$\Gamma^{(2,1)}$ produces a renormalization of the would be
Lagrangian for the $R$ wall, since it correspond to terms which are
proportional to the square of $\eta$ and of its time derivative. They are
local in time, and therefore they will not contribute to any dissipative
effect (which necessarily correspond to non-analyticities in the frequency
space).

Let us then extract from the second order term its imaginary part, which is
related to the total probability of pair creation $P$. Indeed,
the vacuum persistence probability is related to $\Gamma$ by:
\begin{equation}\label{eq:in-out}
	|\braket{0_\text{out}|0_\text{in}}|^2 = e^{-2 \, \text{Im} \, \Gamma}
	\simeq 1 - P,
\end{equation}
where the last equality is valid for $\text{Im} \, \Gamma \ll 1$. This is
essentially the equation for probability conservation, where $P = 2 \,
\text{Im} \, \Gamma$ is the probability of the transition of the vacuum to
a state with a non-vanishing  particle content. Because the first
non-trivial process is the creation of a particle and anti-particle pair,
by computing $\text{Im}\,\Gamma$ we should obtain the pair-production
probability.

In $\Gamma^{(2,2)}$, for bag boundary conditions, and evaluating the trace
over the spatial coordinates
\begin{equation}
	\Gamma^{(2,2)} = - 2i \int_{x^0,y^0} \, \eta(x^0)\eta(y^0)
	\, \partial_{x_1} \partial_{y_1} \text{tr} \left[S_F(x,y)
	S_F(y,x)\right]\big|_{x_1 = y_1 = a}.
\end{equation}
To evaluate the integrals in the last expression, rather than using the
time Fourier transforms, and evaluate the convolution of the propagators,
we take into account that we are interested in a process whereby real
particles are created. Therefore,  the flux of energy will have a definite
sense in the diagram and, in the spirit of the `largest time
equation'~\cite{Veltman:1994wz}, we have found it convenient to use the following
decomposition of the propagator in terms of positive- and negative-energy
projectors:
\begin{equation}
	\begin{aligned}
		S_F(x,y) = \sum_n \big[ &\theta(x_0 - y_0) \, e^{-E_n(x_0 - y_0)} \,
		\mathcal{P}_n^+(x_1,y_1) \\
		- \, &\theta(y_0 - x_0) \, e^{-E_n(y_0 - x_0)} \, \mathcal{P}_n^-(x_1,y_1) \big].
	\end{aligned}
\end{equation}
The energy projectors are written in terms of the solutions of the Dirac
equation with bag boundary conditions:
\begin{equation}
	\mathcal{P}_n^+(x_1,y_1) = u_n(x^1)\bar{u}_n(y^1), \quad
	\mathcal{P}_n^-(x_1,y_1) = v_n(x^1)\bar{v}_n(y^1).
\end{equation}
Evaluating the effective action with the previous representation for the
propagator, we obtain the expression:
\begin{equation}
	\begin{aligned}
		\Gamma^{(2,2)} = & \, 2i \, \sum_{n,l}
		\big|\left(\bar{u}_n(x^1)v_l(x^1)\right)^\prime\big|^2\big|_{x^1 = a}
		\int_{x^0,y^0} \eta(x^0) \eta(y^0)\\
		&[\theta(x^0 - y^0) \, e^{-i(E_n + E_l)(x^0 - y^0)} + \theta(y^0 - x^0)
		\, e^{-i(E_n + E_l)(y^0 - x^0)}].
	\end{aligned}
\end{equation}
Finally, using the integral representation of Heaviside's step function, and expressing
the function $\eta$ in terms of its Fourier transform we get:
\begin{equation}
	\Gamma^{(2,2)} = - 4 \sum_{n,l}
		\big|\left(\bar{u}_n(x^1)v_l(x^1)\right)^\prime\big|^2\big|_{x^1 = a} 
		\int \frac{d\nu}{2\pi}
		\frac{|\tilde{\eta}(\nu)|^2}{\nu - (E_n + E_l) + i\varepsilon}.
\end{equation}
The imaginary part of the last result may be taken in a rather
straightforward way, leading to the result:
\begin{equation}
	P = 2 \, \text{Im} \, \Gamma^{(2,2)}=  4 \sum_{n,l}
	|\tilde{\eta}(E_n + E_l)|^2 \,
	\big|\left(\bar{u}_n(x^1)v_l(x^1)\right)^\prime\big|^2\big|_{x^1 =
	a} \;.
\end{equation}
Namely,
\begin{equation}
	P \;=\; \sum_{n,l} |T_{nl}|^2 \;,
\end{equation}
with $T_{nl}$ as given in (\ref{eq:tfi}); therefore in total agreement with
the results previously obtained.

\section{Conclusions}\label{sec:conc}
In this work, using an $S$-matrix approach, we have evaluated the
fermion pair creation propability for a trembling cavity which enforces bag
boundary conditions on the Dirac field, in $1+1$ dimensions. The results
may be expressed in a rather general form in terms of the eigenenergies of
the static cavity, which in turn correspond to the roots of a
transcendental equation.  In the massless case, results may be written more
explicitly.

We have shown the consistency of those results with the ones stemming from
the imaginary part of the effective action, for the evaluation of which we
have used a shortcut approach.

\section*{Acknowledgements}
The authors thank ANPCyT, CONICET and UNCuyo for financial support.
We thank Dr.~Alessandro Ferreri (J\"ulich), for pinpointing a mistake 
in Eq.~(28) in a previous version of this article, which lead to a wrong expression
for $f(k)$ in Eq.~(36).

\end{document}